\documentclass[a4paper,12pt,leqno]{article}
\usepackage{graphicx}
\usepackage{amsmath}
\usepackage{amssymb}
\usepackage{amstext}
\usepackage{amsxtra}
\usepackage[mathscr]{eucal}

\makeatletter
\@addtoreset{equation}{section}
\renewcommand\theequation
  {\ifnum \c@section>\z@ \thesection.\fi \@arabic\c@equation}
\makeatother

\newcommand{\II}{I\kern -1mm I}
\newcommand{\III}{I\kern -1mm I\kern -1mm I}

\newcommand{\sh}{\kern 0.5mm{\rm sh}}

\title{\bf Confounding of three binary-variables counterfactual model}
\date{}

\author{
\medskip
Jingwei Liu \footnote{Corresponding author. jwliu@buaa.edu.cn (J.W Liu)}, Shuang Hu \\
School of Mathematics and System Sciences,\\
Beihang University,\\
LMIB of the Ministry of Education, \\
Beijing, P.R China, 100191.
}

\begin{document}
\maketitle

\begin{abstract}
Confounding of three binary-variables counterfactual model is discussed in this paper. According to the effect between the control variable and the covariate variable, we investigate three counterfactual models: the control variable is independent of the covariate variable, the control variable has the effect on the covariate variable and the covariate variable affects the control variable. Using the ancillary information based on conditional independence hypotheses, the sufficient conditions to determine whether the covariate variable is an irrelevant factor or a confounder in each counterfactual model are obtained.\\

\end{abstract}

\section{Introduction}

Causal inference has become an important research field in statistics, data mining, epidemiology and machine learning in recent decades (Kleinbaum \textit{et al.}., 1982; Rothman, 1986; Greenland, Robins and Pearl, 1999; Zheng \textit{et al.}, 2001; Geng \textit{et al.}, 2004; Xie and Geng, 2008). Confounding and confounder are two basic concepts for causal inference(Kleinbaum \textit{et al.}., 1982; Greenland and Robins, 1986). Several models have been presented for
causal inference, two of which are the causal diagram model and counterfactual model(Rubin, 1974; Holland, 1986; Geng \textit{et al.}, 2004).

In the presence of confounding bias, the effect of exposure on the rate of a disease can not be assessed correctly. By Greenland and Robins (1986), there are basically two main approaches for assessing confounding and confounder. One approach, which is called 'collapsibility-based', regards confounding bias as arising from difference between stratified measures of association and the corresponding original measure. The other, which is called 'comparability-based', regards confounding bias as arising from the exposed and unexposed populations which are not comparable. The comparability-based approach determines a factor to be a confounder if adjusting for it can reduce confounding bias(Greenland and Robins, 1986; Greenland, Robins and Pearl, 1999). Geng \textit{et al.}.(2002) mainly discussed the confounding of multi-value variables and give the criteria for confounding. Zheng \textit{et al.}.(2002), Liang and Zheng (2003) discussed the identifiability of the causal effect of two kinds of counterfactual models using the independence hypotheses respectively.

Traditionally, a confounding variable(the precise definition of a confounder) is a variable which is a common cause of both the control variable and the response variable (Wunsch, 2007). Whether the covariate variable which is not a common cause of both the control variable and the response variable in three binary-variables counterfactual models must not be a confounder? Kleinbaum \textit{et al.}.(1982), Greenland and Robins(1986) gave a qualitative definition
of confounder: controlling a variable can reduce confounding, then the variable is called a confounder. So the covariate variable, which is not a common cause of both the control variable and the response variable but must affect the response variable, may be a confounder. In this paper, according to the precise definition, one model as shown in Fig1 is discussed: the covariate variable affects the control variable and the response variable at the same time, and the control variable affects the response variable. By the qualitative definition, we investigate other two models: one, as shown in Fig2, is that the control variable has the effect on the covariate variable and the covariate variable affects the response variable; the other, as shown in Fig3, is that the control variable is independent of the covariate variable and the covariate variable affects the response variable. Obviously, the third model is the
special case of the other two models with independence of the control variable and the covariate variable. Then we use the formal definitions of  a confounder and an irrelevant factor in Geng {\em et al.} (2002) and the ancillary information based on conditional independence hypotheses(Zheng \textit{et al.}, 2002; Liang and Zheng, 2003) to discuss the confounding of above-mentioned counterfactual
models.

The rest of the paper is organized as follows: In section 2, we introduce the main notation and definitions. In section 3, confounding of three kinds of three-binary-variables counterfactual models are discussed respectively. Section 4 concludes the paper.

\section{Notation and definitions}

Let $E,D,C$ be binary variables. Let the control variable $E$ be an exposure with the values $e$ and $\bar{e}$ representing "exposed" and "unexposed" respectively. Let the response variable $D$ be an outcome with the values 0 and 1 denoting the presence or absence of a disease, where ${D_{e}}$ is the corresponding response when $E=e$ and ${D_{\bar{e}}}$ is the corresponding response when $E=\bar{e}$ , both of which take values 1 or 0 denoting the presence or absence of a disease. Let $C$ be a covariate variable with possible values 0 or 1.

Many kinds of studies focus on the effects of exposure on the rate of a disease in the exposed population. Let
$P(D_{\bar{e}}=1|E=\bar{e})$ and $P(D_{e}=1|E=e)$ be the proportions of diseased individuals in the unexposed population and the exposed population. Let $P(D_{\bar{e}}=1|E=e)$ be the hypothetical proportion of individuals in the exposed population who would have attacked by the disease even if they had not been exposed. Since $P(D_{\bar{e}}=1|E=e)$ is a hypothetical proportion, the model (Rubin, 1974; Holland, 1986) is a counterfactual model.

In order to identify the casual effect of exposure on response, confounding bias $B$ is defined as the difference between the hypothetical proportion of diseased individuals in the exposed population (Wickramaratneand and Holford, 1987; Holland, 1989), that is
\begin{equation}
B=P(D_{\bar{e}}=1|E=e)-P(D_{\bar{e}}=1|E=\bar{e}).
\end{equation}
If $B=0$, then there is no confounding.

By the common standardization in epidemiology (Miettinen, 1972; Kleinbaum, 1982; Rothman, 1986; Geng \textit{et al.}, 2002), the standardized proportion $P_{\triangle}(D_{\bar{e}}=1|E=\bar{e})$ which is obtained by adjusting the distribution of $C$ in the unexposed population to that in the exposed population is
\begin{equation}
\begin{array}{l}
 P_{\triangle}(D_{\bar{e}}=1|E=\bar{e})\\
=\sum\limits_{k=0}^{1}P(D_{\bar{e}}=1|E=\bar{e},C=k)P(C=k|E=e).
\end{array}
\end{equation}

\textbf{Definition 1}$^{[7]}$. A covariate $C$ is a confounder if
\begin{equation}
|P(D_{\bar{e}}=1|E=e)-P_{\triangle}(D_{\bar{e}}=1|E=\bar{e})|<|B|.
\end{equation}
From the definition, we find that the standardized proportion $P_{\triangle}(D_{\bar{e}}=1|E=\bar{e})$ obtained by adjusting for the irrelevant factor is closer to the hypothetical proportion $P(D_{\bar{e}}=1|E=e)$ than the observed proportion $P(D_{\bar{e}}=1|E=\bar{e})$.

\textbf{Definition 2}$^{[7]}$. A covariate $C$ is an irrelevant factor if
\begin{equation}
P_{\bigtriangleup}(D_{\bar{e}}=1|E=\bar{e})=P(D_{\bar{e}}=1|E=\bar{e}).
\end{equation}

Since the estimation of the hypothetical proportion is still unchanged after being adjusted for an irrelevant factor, we do not need to adjust for it to reduce confounding bias.\\

\textbf{Lemma 1}. If a covariate $C$ is an irrelevant factor, it must not be a confounder. Obviously, the inverse negative proposition is also true.\\

\textbf{Proof}. From the condition that $C$ is an irrelevant factor, we can obtain that
$$P_{\triangle}(D_{\bar{e}}=1|E=\bar{e})=P(D_{\bar{e}}=1|E=\bar{e}).$$

Then,
$$
\begin{array}{l}
|P(D_{\bar{e}}=1|E=e)-P_{\triangle}(D_{\bar{e}}=1|E=\bar{e})|\\
=|P(D_{\bar{e}}=1|E=e)-P(D_{\bar{e}}=1|E=\bar{e})|=|B|.
\end{array}
$$
So, $C$ is not a confounder.

From the condition that $C$ is a confounder, we can obtain
$$
\begin{array}{l}
|P(D_{\bar{e}}=1|E=e)-P_{\triangle}(D_{\bar{e}}=1|E=\bar{e})|\\
<|B|=|P(D_{\bar{e}}=1|E=e)-P(D_{\bar{e}}=1|E=\bar{e})|.
\end{array}
$$
Then,
$$P_{\triangle}(D_{\bar{e}}=1|E=\bar{e})\neq P(D_{\bar{e}}=1|E=\bar{e}).$$
So, $C$ is not an irrelevant factor.

However, the converse proposition of lemma 1 is not always true. For example$^{[7]}$, let a factor $C$ express groups categorized by every 10 years of age, and its values 1, 2, 3 and 4 denote the original age groups 20-29, 30-39, 40-49 and 50-59 years respectively. Suppose that there is no exposure effect, i.e. there are only individuals of type 1 (individual 'doomed') and type 4 (individual immune to disease), and that the joint distribution of
disease, exposure and a factor $C$ is given in Table 1. Here$^{[7]}$

$$
\begin{array}{l}
P(D_{\bar{e}}=1|E=e)=0.52,P(D_{\bar{e}}=1|E=\bar{e})=0.58,\\
|B|=0.06.
\end{array}
$$

When the individuals are regrouped by younger than 50, we can obtain a coarse subpopulation given in Table1.\\

\textmd{\textbf{Table1}}\\
\begin{tabular}{|c|c|c|c|c|}
\hline
       &\multicolumn{4}{c|}{Distribution for the values of $C$}\\
       \cline{2-5}
Type   &\multicolumn{2}{c|}{$C\in\{1,2,3\}$} &\multicolumn{2}{c|}{$C\in\{4\}$}\\
\hline {} &$E=e$&$E=\bar{e}$&$E=e$&$E=\bar{e}$\\\cline{1-5}
1('doomed') &133&122&23&52\\
\cline{1-5}
4('immune') &117&78&27&48\\
\cline{1-5}
Total       &250&200&50&100\\
\cline{1-5} \hline
\end{tabular}\\

Then,

$$
\begin{array}{ll}
&P_{\bigtriangleup}(D_{\bar{e}}=1|E=\bar{e})=P(D_{\bar{e}}=1|E=\bar{e})\\
&=\displaystyle
\frac{122}{200}\times\frac{250}{300}+\frac{52}{100}\times\frac{50}{300}=0.595\neq0.58.
\end{array}
$$

So,

$$
\begin{array}{l}
|P(D_{\bar{e}}=1|E=e)-P_{\triangle}(D_{\bar{e}}=1|E=\bar{e})|\\
=|0.595-0.52|=0.075>|B|.
\end{array}
$$

To sum up, $C$ is not a confounder, but it is not an irrelevant factor.

What's more, the negative proposition of lemma 1 is also not always true. For example$^{[7]}$, when the individuals are regrouped by younger than 40 but older than 30, we can obtain a coarse subpopulation given in Table2.\\

\textmd{\textbf{Table2}}\\
\begin{tabular}{|c|c|c|c|c|}
\hline
            &\multicolumn{4}{c|}{Distribution for the values of $C$}\\
            \cline{2-5}
Type        &\multicolumn{2}{c|}{$C\in\{1,3,4\}$} &\multicolumn{2}{c|}{$C\in\{2\}$}\\
\hline {} &$E=e$&$E=\bar{e}$&$E=e$&$E=\bar{e}$\\\cline{1-5}
1('doomed') &46&26&110&148\\
\cline{1-5}
4('immune') &54&24&90&102\\
\cline{1-5}
Total       &100&50&200&250\\
\cline{1-5} \hline
\end{tabular}\\

Then,
$$
\begin{array}{ll}
&P_{\bigtriangleup}(D_{\bar{e}}=1|E=\bar{e})=P(D_{\bar{e}}=1|E=\bar{e})\\
&=\displaystyle
\frac{26}{50}\times\frac{100}{300}+\frac{148}{250}\times\frac{200}{300}=0.568\neq0.58.
\end{array}
$$
So,
$$
\begin{array}{l}
|P(D_{\bar{e}}=1|E=e)-P_{\triangle}(D_{\bar{e}}=1|E=\bar{e})|\\
=|0.568-0.52|=0.048<|B|.
\end{array}
$$
To sum up, $C$ is not an irrelevant factor, but it is not a confounder.

\section{Confounding of counterfactual model}

Considering the confounding of three binary-variables counterfactual
model, there are three counterfactual models as follows:

\vspace*{\baselineskip} \unitlength=1mm
\begin{picture}(60,50)
\put(2,35){\circle{5}} \put(2,41){$E\{e,\bar{e}\}$}
\put(29,35){\vector(-1,0){24}}
\put(4,33){\vector(1,-1){11}} \put(17,20){\circle{5}}
\put(14,14){$D\{0,1\}$} \put(32,35){\circle{5}}
\put(26,41){$C\{0,1\}$} \put(30,33){\vector(-1,-1){11}}
\put(3,3){Fig1.The first model}
\put(52,35){\circle{5}}
\put(52,41){$E\{e,\bar{e}\}$}
\put(55,35){\vector(1,0){24}}
\put(54,33){\vector(1,-1){11}}
\put(67,20){\circle{5}}
\put(64,14){$D\{0,1\}$}
\put(82,35){\circle{5}}
\put(76,41){$C\{0,1\}$}
\put(80,33){\vector(-1,-1){11}}
\put(53,3){Fig2.The second model}
\put(102,35){\circle{5}}
\put(102,41){$E\{e,\bar{e}\}$}
\put(104,33){\vector(1,-1){11}}
\put(117,20){\circle{5}}
\put(114,14){$D\{0,1\}$}
\put(132,35){\circle{5}}
\put(126,41){$C\{0,1\}$}
\put(130,33){\vector(-1,-1){11}}
\put(103,3){Fig3.The third model}
\end{picture}

To discuss whether there be confounding in our considering models,
we use the conditional independence hypotheses as follows as the
ancillary information($\mathscr{H}$):
$$
\begin{array}{l}
(1) E \bot D_{\bar{e}}. \\
(2) E \bot D_{\bar{e}}|C=0. \\
(3) E \bot D_{\bar{e}}|C=1. \\
(4) E \bot C . \\
(5) D_{\bar{e}} \bot C. \\
(6) D_{\bar{e}} \bot C|E=\bar{e}. \\
(7) D_{\bar{e}} \bot C|E=e.
\end{array}
$$

\subsection{The first model}

As shown in Fig1, $C$ has effect on $E$ and $D$ at the same time,
and $C$ affects $E$. In order to calculate simply, suppose:

$$
\begin{array}{l}
P(C=1)=t,\\
P(E=e|C=1)=a_{1},P(E=e|C=0)=a_{0},\\
P(D_{\bar{e}}=1|E=\bar{e},C=j)=b_{j},\\
P(D_{\bar{e}}=1|E=e,C=0)=u_{0},\\
P(D_{\bar{e}}=1|E=e,C=1)=u_{1},
\end{array}
$$

where
$$\bar{t}=1-t,\bar{a_{0}}=1-a_{0},\bar{a_{1}}=1-a_{1},\bar{b_{j}}=1-b_{j}.$$
$t,a_{0},a_{1},b_{j}$ can be observed from original data, but $u_{0},u_{1}$ can not be observed because they are hypothetical proportions.\\

Then,
$$
\begin{array}{l}
P_{\bigtriangleup}(D_{\bar{e}}=1|E=\bar{e}) \\
=\sum\limits_{k=0}^1{P(D_{\bar{e}}=1|E=\bar{e},C=k)P(C=k|E=e)}\\
=\displaystyle\frac{b_{0}a_{0}\bar{t}+b_{1}a_{1}t}{a_{0}\bar{t}+a_{1}t}.
\end{array}
$$

$$
\begin{array}{l}
P(D_{\bar{e}}=1|E=e)\\
=\sum\limits_{k=0}^1{P(D_{\bar{e}}=1|E=e,C=k)P(C=k|E=e)}\\
=\displaystyle\frac{u_{0}a_{0}\bar{t}+u_{1}a_{1}t}{a_{0}\bar{t}+a_{1}t}.
\end{array}
$$

$$
\begin{array}{l}
P(D_{\bar{e}}=1|E=\bar{e})\\
=\sum\limits_{k=0}^1{P(D_{\bar{e}}=1|E=\bar{e},C=k)P(C=k|E=\bar{e})}\\
=\displaystyle\frac{b_{0}\bar{a_{0}}\bar{t}+b_{1}\bar{a_{1}}t}{\bar{a_{0}}\bar{t}+\bar{a_{1}}t}.
\end{array}
$$

So,
$$
\begin{array}{l}
B=P(D_{\bar{e}}=1|E=e)-P(D_{\bar{e}}=1|E=\bar{e})\\
=\displaystyle\frac{b_{0}a_{0}\bar{t}+b_{1}a_{1}t}{a_{0}\bar{t}+a_{1}t}-\frac{b_{0}\bar{a_{0}}\bar{t}+b_{1}\bar{a_{1}}t}{\bar{a_{0}}\bar{t}+\bar{a_{1}}t}.
\end{array}
$$

Now translate each condition of ($\mathscr{H}$) into parameter form:

$$
\begin{array}{l}
(1).  E\perp D_{\bar{e}}\  \ \textit{i.e.} \hspace*{0.2cm}\displaystyle\frac{b_{0}a_{0}\bar{t}+b_{1}a_{1}t}{a_{0}\bar{t}+a_{1}t}=\displaystyle\frac{b_{0}\bar{a_{0}}\bar{t}+b_{1}\bar{a_{1}}t}{\bar{a_{0}}\bar{t}+\bar{a_{1}}t}\\
(2).  E\perp D_{\bar{e}}|C=0\  \ \textit{i.e.} \  \ u_{0}=b_{0}\\
(3).  E\perp D_{\bar{e}}|C=1\  \ \textit{i.e.} \  \ u_{1}=b_{1}\\
(4).  C\perp D_{\bar{e}}\  \ \textit{i.e.} \  \ u_{0}a_{0}+b_{0}\bar{a_{0}}=u_{1}a_{1}+b_{1}\bar{a_{1}}\\
(5).  C\perp D_{\bar{e}}|E=\bar{e}\  \ \textit{i.e.} \  \ b_{0}=b_{1}\\
(6).  C\perp D_{\bar{e}}|E=e\  \ \textit{i.e.} \  \ u_{0}=u_{1}\\
(7).  C\perp E\  \ \textit{i.e.} \  \ a_{0}=a_{1}
\end{array}
$$

\textbf{Theorem 1}. If

$$
\begin{array}{l}
(a) E \bot C \  \ or\\
(b) D_{\bar{e}} \bot C|E =\bar{e} \  \ or\\
(c) D_{\bar{e}} \bot C|E=e,E \bot D_{\bar{e}}|C
\end{array}
$$
The covariate $C$ is an irrelevant factor.\\

\textbf{Proof}.

In order to prove $C$ is an irrelevant factor, we only need to prove

$$\displaystyle\frac{b_{0}a_{0}\bar{t}+b_{1}a_{1}t}{a_{0}\bar{t}+a_{1}t}=\displaystyle\frac{b_{0}\bar{a_{0}}\bar{t}+b_{1}\bar{a_{1}}t}{\bar{a_{0}}\bar{t}+\bar{a_{1}}t}.$$

That is,

$$(b_{0}-b_{1})(a_{0}-a_{1})=0.$$

(a).  From the condition $E \bot C$, we can obtain $$a_{0}=a_{1}.$$
So,
$$P_{\bigtriangleup}(D_{\bar{e}}=1|E=\bar{e})=P(D_{\bar{e}}=1|E=\bar{e}).$$

(b).  From the condition $D_{\bar{e}} \bot C|E =\bar{e}$, we can
obtain
$$b_{0}=b_{1}.$$
So,
$$P_{\bigtriangleup}(D_{\bar{e}}=1|E=\bar{e})=P(D_{\bar{e}}=1|E=\bar{e}).$$

(c).  From the condition $E \bot D_{\bar{e}}|C$, we can obtain
$$
\begin{array}{l}
E\perp D_{\bar{e}}|C=0\  \ \textit{i.e.} \  \ u_{0}=b_{0},\\
E\perp D_{\bar{e}}|C=1\  \ \textit{i.e.} \  \ u_{1}=b_{1}.
\end{array}
$$

Furthermore,
$$C\perp D_{\bar{e}}|E=e\  \ \textit{i.e.} \  \ u_{0}=u_{1}.$$

Thus,$$b_{0}=b_{1}.$$ So,
$$P_{\bigtriangleup}(D_{\bar{e}}=1|E=\bar{e})=P(D_{\bar{e}}=1|E=\bar{e}).$$

\textbf{Theorem 2}. If

$$
\begin{array}{l}
(a) E \bot D_{\bar{e}} \  \ or\\[0.2em]
(b) C \bot D_{\bar{e}}|E=\bar{e},E \bot D_{\bar{e}}|C \  \ or\\
(c) E \bot D_{\bar{e}}|C=0,C \bot D_{\bar{e}}|E \  \ or\\
(d) E \bot D_{\bar{e}}|C=1,C \bot D_{\bar{e}}|E \  \ or\\
(e) E \bot D_{\bar{e}}|C,E \bot C
\end{array}
$$

There is no confounding.\\

\textbf{Proof}.

(a).From the condition  $E\bot D_{\bar{e}}$, we can obtain

$$\displaystyle\frac{b_{0}a_{0}\bar{t}+b_{1}a_{1}t}{a_{0}\bar{t}+a_{1}t}=\displaystyle\frac{b_{0}\bar{a_{0}}\bar{t}+b_{1}\bar{a_{1}}t}{\bar{a_{0}}\bar{t}+\bar{a_{1}}t}.$$

Then,
$$B=\displaystyle\frac{b_{0}a_{0}\bar{t}+b_{1}a_{1}t}{a_{0}\bar{t}+a_{1}t}=\displaystyle\frac{b_{0}\bar{a_{0}}\bar{t}+b_{1}\bar{a_{1}}t}{\bar{a_{0}}\bar{t}+\bar{a_{1}}t}=0.$$

(b).From the condition that $E \bot D_{\bar{e}}|C$, we can obtain
$$
\begin{array}{l}
E\perp D_{\bar{e}}|C=0$\  \ \textit{i.e.} \  \ $u_{0}=b_{0},\\
E\perp D_{\bar{e}}|C=1$\  \ \textit{i.e.} \  \ $u_{1}=b_{1}.
\end{array}
$$

Furthermore,
$$C\perp D_{\bar{e}}|E=\bar{e}\  \ \textit{i.e.} \  \ b_{0}=b_{1}.$$
Then,
$$B=\displaystyle\frac{b_{0}a_{0}\bar{t}+b_{1}a_{1}t}{a_{0}\bar{t}+a_{1}t}=\displaystyle\frac{b_{0}\bar{a_{0}}\bar{t}+b_{1}\bar{a_{1}}t}{\bar{a_{0}}\bar{t}+\bar{a_{1}}t}=0.$$

(c).From the conditions that $C \bot D_{\bar{e}}|E$, we can obtain
$$
\begin{array}{l}
C\perp D_{\bar{e}}|E=\bar{e}$\  \ \textit{i.e.} \  \ $b_{0}=b_{1},\\
C\perp D_{\bar{e}}|E=e$\  \ \textit{i.e.} \  \ $u_{0}=u_{1}.
\end{array}
$$

Furthermore,
$$E\perp D_{\bar{e}}|C=0\  \ \textit{i.e.} \  \ u_{0}=b_{0}.$$

So,
$$B=\displaystyle\frac{b_{0}a_{0}\bar{t}+b_{1}a_{1}t}{a_{0}\bar{t}+a_{1}t}=\displaystyle\frac{b_{0}\bar{a_{0}}\bar{t}+b_{1}\bar{a_{1}}t}{\bar{a_{0}}\bar{t}+\bar{a_{1}}t}=u_{0}-b_{0}=0.$$

(d).From the conditions that $C \bot D_{\bar{e}}|E$, we can obtain
$$
\begin{array}{l}
C\perp D_{\bar{e}}|E=\bar{e}\  \ \textit{i.e.} \  \ b_{0}=b_{1},\\
C\perp D_{\bar{e}}|E=e\  \ \textit{i.e.} \  \ u_{0}=u_{1}.
\end{array}
$$

Furthermore,
$$E\perp D_{\bar{e}}|C=1\hspace*{0.2cm}\textit{i.e.}\hspace*{0.2cm}u_{1}=b_{1}.$$

So,
$$B=\displaystyle\frac{b_{0}a_{0}\bar{t}+b_{1}a_{1}t}{a_{0}\bar{t}+a_{1}t}=\displaystyle\frac{b_{0}\bar{a_{0}}\bar{t}+b_{1}\bar{a_{1}}t}{\bar{a_{0}}\bar{t}+\bar{a_{1}}t}=u_{1}-b_{1}=0.$$

(e).From the condition that $E \bot D_{\bar{e}}|C$, we can obtain
$$
\begin{array}{l}
E\perp D_{\bar{e}}|C=0$\hspace*{0.2cm}\textit{i.e.}\hspace*{0.2cm}$u_{0}=b_{0},\\
E\perp D_{\bar{e}}|C=1$\  \ \textit{i.e.} \  \ $u_{1}=b_{1}.
\end{array}
$$

Furthermore,
$$C\perp E\  \ \textit{i.e.} \  \ a_{0}=a_{1}.$$
So,
$$B=\displaystyle\frac{b_{0}a_{0}\bar{t}+b_{1}a_{1}t}{a_{0}\bar{t}+a_{1}t}=\displaystyle\frac{b_{0}\bar{a_{0}}\bar{t}+b_{1}\bar{a_{1}}t}{\bar{a_{0}}\bar{t}+\bar{a_{1}}t}=0.$$

\subsection{The second model}

As shown in Fig2, $E$ and $C$ has effect on $D$ at the same time,
and $E$ affects $C$. In order to calculate simply, suppose:

$$
\begin{array}{l}
P(E=e)=a,\\
P(C=1|E=e)=c_{1},P(C=1|E=\bar{e})=c_{0},\\
P(D_{\bar{e}}=1|E=\bar{e},C=j)=b_{j},\\
P(D_{\bar{e}}=1|E=e,C=0)=u_{0},\\
P(D_{\bar{e}}=1|E=e,C=1)=u_{1},
\end{array}
$$

where
$$\bar{a}=1-a,\bar{c_{0}}=1-c_{0},\bar{c_{1}}=1-c_{1},\bar{b_{j}}=1-b_{j}.$$
$a,b_{j}$ can be observed from original data, but $c_{0},c_{1},u_{0},u_{1}$ can not be observed because they are hypothetical proportions.\\

Then,
$$
\begin{array}{l}
P_{\bigtriangleup}(D_{\bar{e}}=1|E=\bar{e})\\
=\sum\limits_{k=0}^1{P(D_{\bar{e}}=1|E=\bar{e},C=k)P(C=k|E=e)}\\
=b_{0}\bar{c_{1}}+b_{1}c_{1}.
\end{array}
$$

$$
\begin{array}{l}
P(D_{\bar{e}}=1|E=e)\\
=\sum\limits_{k=0}^1{P(D_{\bar{e}}=1|E=e,C=k)P(C=k|E=e)}\\
=u_{0}\bar{c_{1}}+u_{1}c_{1}.
\end{array}
$$

$$
\begin{array}{l}
P(D_{\bar{e}}=1|E=\bar{e})\\
=\sum\limits_{k=0}^1{P(D_{\bar{e}}=1|E=\bar{e},C=k)P(C=k|E=\bar{e})}\\
=b_{0}\bar{c_{0}}+b_{1}c_{0}.
\end{array}
$$

So,
$$
\begin{array}{l}
B=P(D_{\bar{e}}=1|E=e)-P(D_{\bar{e}}=1|E=\bar{e})\\
=u_{0}\bar{c_{1}}+u_{1}c_{1}-b_{0}\bar{c_{0}}-b_{1}c_{0}.
\end{array}
$$

Now translate each condition of ($\mathscr{H}$) into parameter form:

$$
\begin{array}{l}
(1).  E\perp D_{\bar{e}}\  \ \textit{i.e.} \  \ u_{0}\bar{c_{1}}+u_{1}c_{1}=b_{0}\bar{c_{0}}+b_{1}c_{0}\\
(2).  E\perp D_{\bar{e}}|C=0\  \ \textit{i.e.} \  \ u_{0}=b_{0}\\
(3).  E\perp D_{\bar{e}}|C=1\  \ \textit{i.e.} \  \ u_{1}=b_{1}\\
(4).  C\perp D_{\bar{e}}\  \ \textit{i.e.} \  \ \displaystyle\frac{u_{0}\bar{c_{0}}\bar{a}+b_{0}{c_{1}}a}{\bar{c_{1}}a+\bar{c_{0}}\bar{a}}=\displaystyle\frac{u_{1}\bar{c_{1}}a+b_{1}{c_{0}}\bar{a}}{c_{0}\bar{a}+c_{1}a}\\
(5).  C\perp D_{\bar{e}}|E=\bar{e}\  \ \textit{i.e.} \  \ b_{0}=b_{1}\\
(6).  C\perp D_{\bar{e}}|E=e\  \ \textit{i.e.} \  \ u_{0}=u_{1}\\
(7).  C\perp E\  \ \textit{i.e.}\  \ c_{0}=c_{1}
\end{array}
$$

\textbf{Theorem 3}. If

$$
\begin{array}{l}
(a) E \bot C \  \ or\\
(b) D_{\bar{e}} \bot C|E =\bar{e} \  \ or\\
(c) D_{\bar{e}} \bot C|E=e,E \bot D_{\bar{e}}|C
\end{array}
$$
The covariate $C$ is an irrelevant factor.\\

\textbf{Proof}.

(a).  From the condition $E\bot C$, we can obtain
$$c_{0}=c_{1}.$$
So,
$$
\begin{array}{l}
P_{\bigtriangleup}(D_{\bar{e}}=1|E=\bar{e})=b_{0}\bar{c_{1}}+b_{1}c_{1}=b_{0}\bar{c_{0}}+b_{1}c_{0}\\
=P(D_{\bar{e}}=1|E=\bar{e}).
\end{array}
$$

(b).  From the condition $D_{\bar{e}} \bot C|E =\bar{e}$, we can
obtain
$$b_{0}=b_{1}.$$
So,
$$
\begin{array}{l}
P_{\bigtriangleup}(D_{\bar{e}}=1|E=\bar{e})=b_{0}\bar{c_{1}}+b_{1}c_{1}=b_{0}(\bar{c_{1}}+c_{1})\\
=b_{0}(\bar{c_{0}}+c_{0})=b_{0}\bar{c_{0}}+b_{1}c_{0}=P(D_{\bar{e}}=1|E=\bar{e}).
\end{array}
$$

(c).  From the condition $D_{\bar{e}} \bot C|E=e$, we can obtain
$$
\begin{array}{l}
E\perp D_{\bar{e}}|C=0\  \ \textit{i.e.} \  \ b_{0}=b_{1},\\
E\perp D_{\bar{e}}|C=1\  \ \textit{i.e.}\  \ u_{0}=u_{1}.
\end{array}
$$

Furthermore,
$$D_{\bar{e}}\bot
C|E=e\  \ \textit{i.e.} \  \ b_{0}=b_{1}.$$

So,
$$P_{\bigtriangleup}(D_{\bar{e}}=1|E=\bar{e})=P(D_{\bar{e}}=1|E=\bar{e}).$$

\textbf{Theorem 4}. If
$$
\begin{array}{l}
(a) E \bot D_{\bar{e}} \  \ or\\
(b) E \bot D_{\bar{e}}|C=0,C \bot D_{\bar{e}}|E \  \ or\\
(c) E \bot D_{\bar{e}}|C=1,C \bot D_{\bar{e}}|E \  \ or\\
(d) E \bot D_{\bar{e}}|C,E \bot C
\end{array}
$$
There is no confounding.\\

\textbf{Proof}.

(a). From the condition that $E\bot D_{\bar{e}}$, we
can obtain
$$u_{0}\bar{c_{1}}+u_{1}c_{1}=b_{0}\bar{c_{0}}+b_{1}c_{0}.$$
Then,
$$B=u_{0}\bar{t}+u_{1}t-b_{0}\bar{t}-b_{1}t=0.$$

(b). From the conditions that $C \bot D_{\bar{e}}|E$, we can obtain
$$
\begin{array}{l}
C\perp D_{\bar{e}}|E=\bar{e}\  \ \textit{i.e.} \  \ b_{0}=b_{1},\\
C\perp D_{\bar{e}}|E=e\  \ \textit{i.e.} \  \ u_{0}=u_{1}.
\end{array}
$$

Furthermore,
$$E\perp D_{\bar{e}}|C=0\  \ \textit{i.e.} \  \ u_{0}=b_{0}.$$

So,
$$B=u_{0}\bar{c_{1}}+u_{1}c_{1}-b_{0}\bar{c_{0}}-b_{1}c_{0}=u_{0}-b_{0}=0.$$

(c). From the conditions that $C \bot D_{\bar{e}}|E$, we can obtain
$$
\begin{array}{l}
C\perp D_{\bar{e}}|E=\bar{e}\  \ \textit{i.e.} \  \ b_{0}=b_{1},\\
C\perp D_{\bar{e}}|E=e\  \ \textit{i.e.} \  \ u_{0}=u_{1}.
\end{array}
$$

Furthermore,
$$E\perp D_{\bar{e}}|C=1\  \ \textit{i.e.}\  \ u_{1}=b_{1}.$$

So,
$$B=u_{0}\bar{c_{1}}+u_{1}c_{1}-b_{0}\bar{c_{0}}-b_{1}c_{0}=u_{1}-b_{1}=0.$$

(d). From the condition that $E \bot D_{\bar{e}}|C$, we can obtain
$$
\begin{array}{l}
E\perp D_{\bar{e}}|C=0\  \ \textit{i.e.}\  \ u_{0}=b_{0},\\
E\perp D_{\bar{e}}|C=1\  \ \textit{i.e.}\  \ u_{1}=b_{1}.
\end{array}
$$

Furthermore,
$$C\perp E\  \ \textit{i.e.} \  \ c_{0}=c_{1}.$$

So,
$$
\begin{array}{l}
B=u_{0}\bar{t}+u_{1}t-b_{0}\bar{t}-b_{1}t\\
=(u_{0}-b_{0})\bar{c_{0}}+(u_{1}t-b_{1})c_{0}=0.
\end{array}
$$

\subsection{The third model}

As shown in Fig3, both of $E$ and $C$ have effects on $D$, and $E
\bot C$. In order to calculate simply, suppose:
$$
\begin{array}{l}
P(E=e)=a,P(C=1)=t,\\
P(D_{\bar{e}}=1|E=\bar{e},C=j)=b_{j},\\
P(D_{\bar{e}}=1|E=e,C=0)=u_{0},\\
P(D_{\bar{e}}=1|E=e,C=1)=u_{1},
\end{array}
$$
where
$$\bar{a}=1-a,\bar{t}=1-t,\bar{b_{j}}=1-b_{j}.$$
$a,t,b_{j}$ can be observed from original data, but $u_{0},u_{1}$ can not be observed because they are hypothetical proportions.\\

Then,
$$
\begin{array}{l}
P_{\bigtriangleup}(D_{\bar{e}}=1|E=\bar{e})\\
=\sum\limits_{k=0}^1{P(D_{\bar{e}}=1|E=\bar{e},C=k)P(C=k|E=e)}\\
=b_{0}\bar{t}+b_{1}t.
\end{array}
$$

$$
\begin{array}{l}
P(D_{\bar{e}}=1|E=e)\\
=\sum\limits_{k=0}^1{P(D_{\bar{e}}=1|E=e,C=k)P(C=k|E=e)}\\
=u_{0}\bar{t}+u_{1}t.
\end{array}
$$

$$
\begin{array}{l}
P(D_{\bar{e}}=1|E=\bar{e})\\
=\sum\limits_{k=0}^1{P(D_{\bar{e}}=1|E=\bar{e},C=k)P(C=k|E=\bar{e})}\\
=b_{0}\bar{t}+b_{1}t.
\end{array}
$$

So,
$$
\begin{array}{l}
B=P(D_{\bar{e}}=1|E=e)-P(D_{\bar{e}}=1|E=\bar{e})\\
=u_{0}\bar{t}+u_{1}t-b_{0}\bar{t}-b_{1}t.
\end{array}
$$
Of course,
$$P_{\bigtriangleup}(D_{\bar{e}}=1|E=\bar{e})=P(D_{\bar{e}}=1|E=\bar{e}),$$
so the covariate $C$ is an irrelevant factor, but not a confounder and cannot reduce confounding.\\
Now translate each condition of ($\mathscr{H}$) into parameter form:

$$
\begin{array}{l}
(1).  E\perp D_{\bar{e}}\  \ \textit{i.e.}\  \ u_{0}\bar{t}+u_{1}t=b_{0}\bar{t}+b_{1}t\\
(2).  E\perp D_{\bar{e}}|C=0\  \ \textit{i.e.}\  \ u_{0}=b_{0}\\
(3).  E\perp D_{\bar{e}}|C=1\  \ \textit{i.e.} \  \ u_{1}=b_{1}\\
(4).  C\perp D_{\bar{e}}\  \ \textit{i.e.} \  \ b_{0}\bar{a}+u_{0}a=b_{1}\bar{a}+u_{1}a\\
(5).  C\perp D_{\bar{e}}|E=\bar{e}\  \ \textit{i.e.} \  \ b_{0}=b_{1}\\
(6).  C\perp D_{\bar{e}}|E=e\ \ \textit{i.e.} \  \ u_{0}=u_{1}
\end{array}
$$

\textbf{Theorem 5}. If
$$
\begin{array}{l}
(a) E \bot D_{\bar{e}} \  \ or\\
(b) E \bot D_{\bar{e}}|C=0,E \bot D_{\bar{e}}|C=1 \  \ or\\
(c) D_{\bar{e}}\bot C|E,E \bot D_{\bar{e}}|C=0 \  \ or\\
(d) D_{\bar{e}}\bot C|E,E \bot D_{\bar{e}}|C=1
\end{array}
$$
There is no confounding.\\

\textbf{Proof}.

(a).  From the condition that $E \bot D_{\bar{e}}$, we
can obtain
$$u_{0}\bar{t}+u_{1}t=b_{0}\bar{t}+b_{1}t.$$
Then,
$$B=u_{0}\bar{t}+u_{1}t-b_{0}\bar{t}-b_{1}t=0.$$

(b).  From the conditions that $$E\bot D_{\bar{e}}|C=0\
 \ and \  \ E\bot D_{\bar{e}}|C=1,$$ we can obtain
$$u_{0}=b_{0}\  \ and \  \ u_{1}=b_{1}.$$
So,
$$B=u_{0}\bar{t}+u_{1}t-b_{0}\bar{t}-b_{1}t=0.$$

(c).  From the condition that $D_{\bar{e}}\bot C|E$, we can obtain
$$
\begin{array}{l}
D_{\bar{e}}\bot C|E=\bar{e}\  \ \textit{i.e.} \  \ b_{0}=b_{1},\\
C\perp D_{\bar{e}}|E=e\  \ \textit{i.e.} \  \ u_{0}=u_{1}.
\end{array}
$$

Furthermore,
$$E\perp D_{\bar{e}}|C=0 \  \ \textit{i.e.} \  \ u_{0}=b_{0}.$$

So,
$$B=u_{0}\bar{t}+u_{1}t-b_{0}\bar{t}-b_{1}t=u_{0}-b_{0}=0.$$

(d).  The proof is similar to (c).

\section{Conclusion}

Using the formal definitions of a confounder and an irrelevant factor, we discuss the confounding of three kinds of three binary-variables counterfactual models in epidemiology studies and statistics. The sufficient conditions of non-confounding and whether the covariate is an irrelevant factor or a confounder are discussed.
The future work will extend the three variables counterfactual model to multi-variable counterfactual model.

\section{Acknowledgements}
This work is partially supported by National Natural Science Foundation of China for Youths (NSFC:10801019).

\end{document}